# A fault detection scheme for PV modules in large scale PV stations with complex installation conditions

Qianni Cao, Chen Shen, *Senior Member*, *IEEE*, Mengshuo Jia, *Student Member*

*Abstract*—Faults in photovoltaic (PV) systems can seriously affect the efficiency, energy yield as well as the security of the entire PV plant, if not detected and corrected quickly. Therefore, fault diagnosis of PV arrays is indispensable for improving the reliability, efficiency, productivity and safety of PV power stations. Instead of conventional thresholding methods and artificial intelligent (AI) machine learning approaches, an innovative Gaussian Mixture Model (GMM) based fault detection approach is proposed in this work. This approach combines the superiority of GMM in modeling stochastic power outputs of PV modules and the flexibility and simplicity of Sandia PV Array Performance Model (SPAM) to accurately detect under-performing modules. Firstly, GMM is proposed to represent the probabilistic distribution functions (PDF) of different PV modules' power outputs, and the parameter sets of which are obtained by Expectation Maximization algorithm. Secondly, a simplified explicit expression for output power of PV modules, which highlights the influences of tilt and azimuth angles, is deduced based on the SPAM. Then, an orientation independent vector *C* is developed to eliminate the probability distribution differences of power outputs caused by varying azimuth angles and tilt angles. Jensen-Shannon (JS) divergence, which captures the differences between probability density of *C* of each PV module, are generated and used as fault indicators. Simulation data acquired from the original SPAM are used to assess the performance of the proposed approach. Results show that the proposed approach successfully detects faults in PV systems. This work is especially suitable for the PV modules that have different installation parameters such as azimuth angles and tilt angles, and it does not require the use of irradiance or temperature sensors.

## I. INTRODUCTION

### A. The State of the Art

SEVERAL fault detection techniques for PV systems have been proposed in literatures and most of them are data-driven. These methods can be classified into two main categories: process-history-based approaches and model-based approaches. In application of artificial intelligence techniques and machine learning methods, process-history-based approaches obtain implicit empirical models with analysis of available data. Paper [1] proposed a fault detection model based on artificial neutral network(ANN) to detect possible solar module abnormalities. Paper [2] demonstrated an ANN technique to estimate the output photovoltaic current and voltage under different working conditions. It requires solar irradiance, PV module's temperature, PV array's current and voltage as inputs to identify the different types of faults. An approach combining Support Vector Machine (SVM) and k-Nearest Neighbor(K-NN) was developed in paper [3]. It is able to detect and locate short circuits, bypass diode faults and blocking diode faults with high accuracy. In [4], the authors presented an Extension Neural Network (ENN) fault diagnosis method to detect totally 9 investigated faults. The technique adopts the solar irradiation, module temperature, open circuit voltage of the PV array and power, current, voltage at Maximum Power Point (MPP) during operation as inputs. However, process-history-based methods require the availability of a relevant dataset which describes both healthy and faulty operating conditions of a PV system. Moreover, they lack the ability to detect new faults.

On the other hand, model-based approaches compare simulation outputs with measured values to make a power loss analysis. Anomaly is declared when large differences are detected. Paper [5] proposed a simple faults detection method by comparing the simulation parameters with those measured. This method requires temperature and irradiance sensors and several power meters. Paper [6] used a Kalman filter to predict the power output of a PV system. In paper [7], an approach combining one-diode model with an exponentially weighted moving average (EWMA) control chart was demonstrated. It captures the difference between the measurements and the predictions of the one-diode model and applies the EWMA monitoring chart to identify the type of fault. However, an error in simulation results may emerge due to inaccuracies of PV performance models and auxiliary equations that translate model parameters to operating conditions. Moreover, both irradiance and solar module temperature measurements are needed for such approaches. Extra sensors required to provide precise measurements can be a significant extra cost to the PV installation.

## B. Motivation and Contributions

In this paper, a method based on Gaussian Mixture Model (GMM) is developed to detect faults in PV systems, especially for those with complex installation conditions, i.e. PV systems installed with varying azimuth angles and tilt angles. The power output of each module is considered as a random variable that follows GMM. The merit of the method is that it does not need the measurement of irradiance and air temperature. It is based on the comparison of the power produced by different modules. In this paper, we suppose that most PV modules operate under normal condition. When failure occurs in a PV module, its probability distribution (PDF) of power output will diverge from the majority. However, differences between PDF of power output of modules are caused by either abnormal conditions or by complex installation conditions. Therefore, the difference due to varying installation parameters should be eliminated, and a novel azimuth and tilt independent random variable should be developed. For modules installed in the same area and of the same specification, we assume that their PDF of the variable are similar. However, if a module has a very different PDF of the variable from the majority, it should be regarded as faulty. Moreover, in our approach, severity of each failure can also be revealed, alerting the users which failures are more severe to solve. None of the approaches presented so far used probabilistic analysis to monitor the power loss and reveal abnormalities.

The paper is organized as follows. Section 2 develops an explicit expression, which highlights the influences of tilt and azimuth angles for power produced by modules. And a both azimuth and tilt independent vector $C$ is then developed. Section 3 introduces the use of $C$ to detect faults. Section 4 applies the proposed fault-detection method to simulation models based on the Sandia PV Array Performance Model (SPAM). Finally, Section 5 concludes with contributions and suggestions for future research directions.

## II. PV Performance Model

### A. Linearized PV Module Model

Numerous models have been reported in the literature that model energy production of PV cells. The one-diode model (ODM) is the most commonly used model to predict energy production from PV cells [8]. Nevertheless, in our method, adopting ODM as PV model may lead to a difficult solution procedure and take a long time for solving, as it finds the maximum power (P=IV) by iteratively solving the current-voltage relation.

In our approach, Sandia PV Array Performance Model (SPAM) [9] is adopted to obtain an explicit expression for power produced by each module, which has been proved to be widely applicable for the photovoltaic industry by detailed tests. In this paper, we suppose that the module always operates at MPP; the DC/DC converter is always extracting the maximum power from the module.

In order to effectively eliminate the output power differences caused by varying azimuth and tilt angles, our approach linearizes the effect of effective irradiance $E_e$ (to which the cells respond) on the module and deduces an expression of output power $P_{mp}$ based on the SPAM, as shown in (1)-(6). Particularly, they stress the influence of tilt and azimuth angles. Tests in section 4 demonstrate that our simplification is reasonable and it brings no negative impact on failure detection of PV modules.

$$P_{mp} = \left[I_{mpo}\left(1 + \gamma_{mp}(T_c - 25)\right)/G_0\right] * V_{mp}C_oOE_e \quad (1)$$

$$E_e = A_1 cos\beta_s + A_2 sin\beta_s cos\gamma_s + A_3 sin\beta_s sin\gamma_s + A_4 \quad (2)$$

where

$$A_1 = DNI \cdot cosZ + (\frac{1}{2}DHI - \frac{1}{2}\rho \cdot GHI) \quad (3)$$

$$A_2 = DNI \cdot sinZcos\gamma \quad (4)$$

$$A_3 = DNI \cdot sinZsin\gamma \quad (5)$$

$$A_4 = \frac{1}{2}DHI + \frac{1}{2}\rho \cdot GHI \quad (6)$$

For further condensation of (1), it can be developed as:

$$P_{mp} = C_1 cos\beta_s + C_2 sin\beta_s cos\gamma_s + C_3 sin\beta_s sin\gamma_s + C_4 \quad (7)$$

where

$$\begin{aligned}\boldsymbol{C} = (C_1, \ C_2, \ C_3, \ C_4)^T = (A_1, \ A_2, \ A_3, \ A_4)^T \\ * \left[I_{mpo}\left(1 + \gamma_{mp}(T_c - 25)\right)/G_0\right]V_{mp}C_oO\end{aligned} \quad (8)$$

Parameters in our linearized model are summarized in TABLE I.

TABLE I
Sandia Array Performance Model Parameter Definition

| Symbol | Description | Type |
|---|---|---|
| DHI | The total solar irradiance on a surface parallel to the ground (horizontal) | meteorological condition of irradiance |
| DNI | Beam normal | |
| GHI | The total solar irradiance on a surface parallel to the ground (horizontal) | |
| Z | Sun zenith angle (deg) | Sun position |
| $\gamma$ | Sun azimuth angle (deg) | |
| $\rho$ | Albedo(ground reflectance) | |
| $I_{mpo}$ | Reference Max Power Current(I) | Inherent parameter |
| $G_o$ | Reference power(W/$m^2$) | |
| $\gamma_{mp}$ | temperature coefficient of the $P_{mpp}$ | |
| $\beta_s$ | Surface tilt angle (deg) | Parameter varying from module to module |
| $\gamma_s$ | Surface azimuth angle (deg) | |
| A | Random vector related to irradiance | Random vector |
| C | A random vector, $\boldsymbol{C} = (C_1, C_2, C_3, C_4)$ to evaluate if any of the modules are underperforming | |

### B. Simulation and Analysis of PV Module Typical Faults

Given in TABLE II, four typical faults or abnormal conditions are simulated by decreasing related parameters as adopted in [10].

Particularly, in the aforementioned linearized model, voltage at MPP is supposed to be independent of orientation, but influenced by module temperature. Since we assume PV modules under normal condition operate at the same temperature, the maximum power voltage $V_{mp}$ of the normal ones is considered uniform as well. Nevertheless, when short-circuits of PV modules happen, $V_{mp}$ of them greatly decreases.

TABLE II
FAULTS RELATED VARIABLES IN SPAM

| Symbol | Description | Fault Type |
|---|---|---|
| $C_o$ | coefficient relating Imp to G, representing power conversion efficiency | Degradation fault |
| $V_{mp}$ | Max power voltage(V) | Short-circuit fault |
| $O$ | Soiling and shading loss | Soiling and shading condition |
| $A$ | Random vector related to irradiance | Open circuit fault |

Besides, the extracted photocurrent of module will be greatly affected by the open-circuit fault[10]. To adapt to our model, a change in probability distribution of $A$ is expected when open-circuit faults occur, since the extracted photo-current is almost linear with the irradiance.

## III. FAULT DETECTION APPROACHES

### A. Modelling uncertainties with GMM

In this paper, the Gaussian Mixture Model (GMM) is proposed to represent the joint distribution of random variables, such as outputs of different PV modules. In data clustering and machine learning, the GMM is known for its high level accuracy in characterizing multiple random variables [11]. Recently, several researchers have applied the GMM technique to power system uncertainty analysis and verified its superiority in modeling stochastic power outputs of renewable energy and loads [12][13]. The advantage of adopting GMM is that it can use the combination of finite Gaussian distributions to well model different types of correlated and non-Gaussian random variables.

More precisely, the joint PDF of random vector $X$ is a convex combination of several multivariate Gaussian PDFs with the parameter set $\Omega = \{\omega_m, \mu_m, \sigma_m; m = 1,2,\dots M\}$:

$$f_X(x) = \sum_{m=1}^{M} \omega_m N_m(x; \mu_m, \sigma_m) \quad (9)$$

$$\sum_{m=1}^{M} \omega_m = 1, \ \omega_m > 0 \quad (10)$$

$$N_m(x; \mu_m, \sigma_m) = \frac{e^{-\frac{1}{2}(x-\mu_m)^T \sigma_m^{-1}(x-\mu_m)}}{(2\pi)^{W/2} \det(\sigma_m)^{1/2}} \quad (11)$$

where $x$ denotes the input random vector with M components; $\omega_m$ is the weight coefficient; $\mu_m$ and $\sigma_m$ are mean vector and covariance matrix of the $m$th Gaussian component. Given historical data samples of $X$, obtaining $\Omega$ is formulated as a parameter maximum likelihood estimation (MLE) problem, and can be effectively solved by the well-known Expectation Maximization algorithm[11]. The calculation process of the $k$th iteration for the $m$th Gaussian component in GMM is given as follows:

$$Q_{j,n}^k = \frac{w_j^{k-1} \mathcal{N}(\varsigma_n; \mu_j^{k-1}, \Sigma_j^{k-1})}{\sum_{j=1}^{J} w_j^{k-1} \mathcal{N}(\varsigma_n; \mu_j^{k-1}, \Sigma_j^{k-1})} \quad (12)$$

$$w_m^k = \frac{1}{N} \sum_{n=1}^{N} Q_{m,n}^k \quad (13)$$

$$\mu_m^k = \frac{\sum_{n=1}^{N} Q_{m,n}^k \varsigma_n}{\sum_{n=1}^{N} Q_{m,n}^k} \quad (14)$$

$$\Sigma_m^k = \frac{\sum_{n=1}^{N} Q_{m,n}^k (\varsigma_n - \mu_m^k)^T (\varsigma_n - \mu_m^k)}{\sum_{n=1}^{N} Q_{m,n}^k} \quad (15)$$

where $\varsigma_n$ is the $n$th historical data sample.

Let random vector $X = (X_1, X_2, \dots X_W)$ denotes the output power of PV modules in a nearby area, where $X_i$ is the random output power of the $i$th module. First, by adopting EM algorithm, joint PDF of random vector $X$ is calculated given historical data samples of the outputs. Then the marginal PDF are calculated to model the output of each module. Due to the strong correlation between the outputs of PV modules in a nearby area, off-diagonal elements of covariance matrix contain the correlation information. Nevertheless, the accuracy of subsequent PV module fault detection will be affected if the parameter set $\Omega$ of each PV module are estimated separately because correlation between the modules are yet to be obtained. In the case of a numerous amount of PV modules, they can be grouped according to their geographical location, and then the joint PDF of each group can be solved to improve the efficiency of fault detection.

If a random vector $X$ is modelled by GMM as shown in (16) ~(20),

$$f_X(x_1, x_2, \dots, x_W) = \sum_{m=1}^{M} \omega_m N_m(x_1, x_2, \dots x_W; \mu_m, \sigma_m) \quad (16)$$

$$N_m(x_1, x_2, \dots x_W; \mu_m, \sigma_m) = \frac{e^{-\frac{1}{2}(x-\mu_m)^T \sigma_m^{-1}(x-\mu_m)}}{(2\pi)^{W/2} \det(\sigma_m)^{1/2}} \quad (17)$$

$$\mu_m = [\mu_m^{x_1}, \mu_m^{x_2}, \dots, \mu_m^{x_W}] \quad (18)$$

$$\sigma_m = \begin{bmatrix} \sigma_m^{x_1 x_1} & \sigma_m^{x_1 x_2} & \cdots & \sigma_m^{x_1 x_W} \\ \sigma_m^{x_2 x_1} & \sigma_m^{x_2 x_2} & \cdots & \sigma_m^{x_2 x_3} \\ \vdots & \vdots & \ddots & \vdots \\ \sigma_m^{x_W x_1} & \sigma_m^{x_M x_2} & \cdots & \sigma_m^{x_W x_W} \end{bmatrix} \quad (19)$$

the marginal PDF of any component in vector $X$, .i.e. $X_1, X_2, \dots, X_W$, are given as

$$f_{X_i}(x_i) = \sum_{m=1}^{M} \omega_m N_m(x_i; \mu_m^{x_i}, \sigma_m^{x_i x_i}) \quad (20)$$

where $i = 1,2,3 \dots, W$.

### B. Least-norm solution of $C$

For each module, the linear transformation from $P_{mp}$ to $C$ is an underdetermined equation, as shown in (21):

$$(C_1, C_2, C_3, C_4)^T = (cos\beta_s, sin\beta_s cos\gamma_s, sin\beta_s sin\gamma_s, 1)^{-1} P_{mp} \quad (21)$$

We calculate pseudo-inverse to obtain the linear transformation from $P_{mp}$ to $C$ and $C^+$ is the least-norm solution of (7).

The pseudo-inverse of nonzero vector $x$ is expressed as

$$x^+ = \frac{x^T}{x^T \cdot x} \quad (22)$$

According to the equation above, pseudo-inverse of $(cos\beta_s, sin\beta_s cos\gamma_s, sin\beta_s sin\gamma_s, 1)$ is given as $\left(\frac{1}{2} cos\beta_s, \frac{1}{2} sin\beta_s cos\gamma_s, \frac{1}{2} sin\beta_s sin\gamma_s, \frac{1}{2}\right)$.

*C. Joint distribution of $\boldsymbol{C}$ when Gaussian distribution adopted*

The linear transformation from $P_{mp}$ to $\boldsymbol{C}$ can be described as $\boldsymbol{Y} = \boldsymbol{PX} + \boldsymbol{Q}$, where $\boldsymbol{X}$ is the column vector of the random quantities $X_1, X_2, \ldots X_W$; $\boldsymbol{Y}$ is the column vector of the random quantities $Y_1, Y_2, \ldots Y_K$. When $\boldsymbol{X}$ is modeled by the multivariate Gaussian distribution $N_m(\boldsymbol{x})$, the output vector $\boldsymbol{Y}$ obeys multivariate Gaussian distribution with mean vector $\boldsymbol{P\mu_m} + \boldsymbol{Q}$ and covariance matrix $\boldsymbol{P\Sigma_m P^T}$. Therefore, the joint PDF of $\boldsymbol{Y}$ is given as follows:

$$N_m(\boldsymbol{y}) = \frac{e^{-\frac{1}{2}(y-P\mu_m-Q)^T(P\Sigma_m P^T)^{-1}(y-P\mu_m-Q)}}{(2\pi)^{K/2}\det(P\Sigma_m P^T)^{1/2}} \quad (23)$$

$$f_Y(\boldsymbol{y}) = \sum_{m=1}^{M} \omega_m N_m(\boldsymbol{y}) \quad (24)$$

For each module, $\boldsymbol{C} = (C_1, C_2, C_3, C_4)^T$ is given as

$$\boldsymbol{C} = (cos\beta_s, sin\beta_s cos\gamma_s, sin\beta_s sin\gamma_s, 1)^{-1} P_{mp} \quad (25)$$

As $P_{mp}$ is modeled by GMM and $\boldsymbol{C}$ is the linear transformation of $P_{mp}$, PDF of $\boldsymbol{C}$ can also be given as a GMM.

However, for the degenerate matrix $\boldsymbol{P\Sigma_m P^T}$, (23) makes no sense since the inverse matrix $(\boldsymbol{P\Sigma_m P^T})^{-1}$ does not exist. For example, if $\boldsymbol{P\Sigma_m P^T}$ is not full rank (e.g., K > W), which means $\det(\boldsymbol{P\Sigma_m P^T}) = 0$, (23) no longer holds. This case is called singular Gaussian distribution or the "degenerate case". The modification of (23) is detailed in [14] and briefly summarized as follows. Suppose that $\boldsymbol{P\Sigma_m P^T}$ has $r (r < K)$ nonzero and $K - r$ zero eigenvalues. Then, the joint of $\boldsymbol{y}$ is given by (26) and (27):

$$N_m(\boldsymbol{y}) = \frac{e^{-\frac{1}{2}\tilde{y}^T R^+ \tilde{y}}}{(2\pi)^{r/2}} \left( \prod_{i=1}^{r} \lambda_i \right)^{-\frac{1}{2}} \prod_{i=r+1}^{K} \delta(\tilde{\boldsymbol{y}}^T U_i) \quad (26)$$

$$\tilde{\boldsymbol{y}} = \boldsymbol{y} - \boldsymbol{P\mu_m} - \boldsymbol{Q} \quad (27)$$

where $\lambda_1, \ldots, \lambda_r$ are $r$ nonzero eigenvalues of $\boldsymbol{P\Sigma_m P^T}$; $\boldsymbol{U_1}, \ldots, \boldsymbol{U_K}$ denote K eigenvectors; $\boldsymbol{R}^+ = \sum_{i=1}^{r} \lambda_i^{-1} \boldsymbol{U_i U_i^T}$ is the pseudoinverse Moore-Penrose matrix of $\boldsymbol{P\Sigma_m P^T}$; each multiplier in (26) with the numbers $r+1, \ldots, K$ tends to a Dirac delta function according to its definition

$$\delta(t) = \lim_{\sigma \to 0}[(2\pi\sigma^2)^{-1/2}\exp\left(-\frac{t^2}{2\sigma^2}\right)] \quad (28)$$

It is easily seen that the vector $\boldsymbol{Y}$, distributed by (26), with probability 1 belongs to the hyperplane family $\tilde{\boldsymbol{y}}^T U_i = 0$, where $i = r+1, \ldots, K$.

*D. Jensen-Shannon divergence between $\boldsymbol{C}$*

In probability theory and statistics, the Jensen-Shannon divergence(JSD) is used as an index to describe the similarity between two probability distributions. It is based on the Kullback-Leibler divergence (KLD), with some notable differences, including that it is symmetric and it always has a finite value. For discrete probability distributions $P$ and $Q$ defined on the same probability space, $\mathcal{X}$, KLD of $Q$ from $P$ is defined as

$$D_{KL}(P \parallel Q) = \sum_{x \in \mathcal{X}} P(x) log\left(\frac{P(x)}{Q(x)}\right) \quad (29)$$

JSD is defined by

$$JSD(P \parallel Q) = \frac{1}{2}D(P \parallel H) + \frac{1}{2}D(Q \parallel H) \quad (30)$$

where $H = \frac{1}{2}(P + Q)$.

By comparing PDF of vector $\boldsymbol{C}$ of each module, faults can be detected.

## IV. RESULTS AND DISCUSSION

In this section, the ability of the proposed scheme to detect the presence of faults in the data is assessed and four case studies involving different types of faults are conducted. In the first case study, it is assumed that the PV system operates in normal conditions. In the second and third case study, degrading variables in common factors in $\boldsymbol{C}$ and $\boldsymbol{A}$ are considered respectively. In the fourth case study, PV system is exposed to various fault conditions given in TABLE II.

Practical data, such as temperature, wind speed and irradiance are collected from Photovoltaic Geographical Information System of European Commission, in which azimuth and tilt angles of PV modules can be set by users. The individual module performance coefficients under normal operating conditions in SPAM are from Sandia National Laboratories. In this section, SPAM is adopted to calculate output power of PV modules in time series, which is consistent with data provided by Photovoltaic Geographical Information System. Hence, we can set tilt angles, azimuth angles and possible degrading variables in (7) according to the different operating conditions in the following four case studies.

In this paper, the hourly data of 2016(totally 8782 records) are collected, among which 4324 records of 2016 are used as training set to estimate the GMM parameters, after excluding data during the nights.

Our tests include 16 PV modules. Most modules operate in normal condition, whereas some of them perform slightly worse than others due to improper settings of tilt or azimuth angles. Nevertheless, some other modules are set at optimal angles, but the power was lower than others due to failures. The purpose of the proposed detection scheme is to differentiate between them and to detect the exact faulty modules. However, our approach cannot detect the exact cause and still needs field data to validate its feasibility in real PV arrays.

*A. Normal operating conditions*

In this case study, azimuth and tilt angles of all PV modules are diverse from each other, while all of them are set to work in the normal condition without any faults. Specifically, module performance coefficients of the 16 modules are set at the same optimal value: $C_0 = 1.0275, O = 1$. The purpose of this case study is to validate the ability of vector $\boldsymbol{C}$ to eliminate the impact of the varying azimuth angle and tilt angle.

For the 16 normal PV modules formulated above, the JS divergence between PDFs of output power and between the vector $\boldsymbol{C}$ are shown in Fig.1. It can be seen that the JS divergence between PDF of $\boldsymbol{C}$ is lower than that of output power, indicating the PDF of $\boldsymbol{C}$ is less influenced by varying azimuth and tilt angles than the PDF of output power.

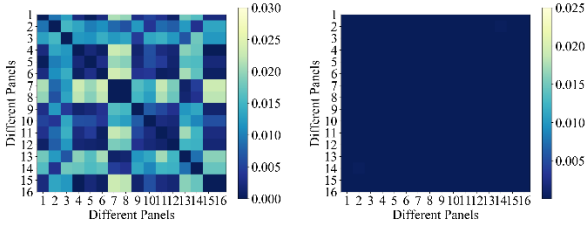

(a) $P_{mp}$          (b) $C$

Fig.1    The JS divergence for 16 PV modules(Case 1)

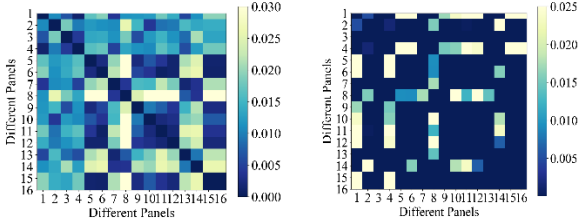

(b) $P_{mp}$          (b) $C$

Fig.2    The JS divergence for 16 PV modules(Case 2)

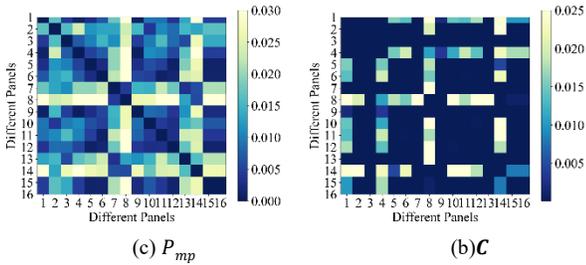

(c) $P_{mp}$          (b) $C$

Fig.3    The JS divergence for 16 PV modules(Case 3)

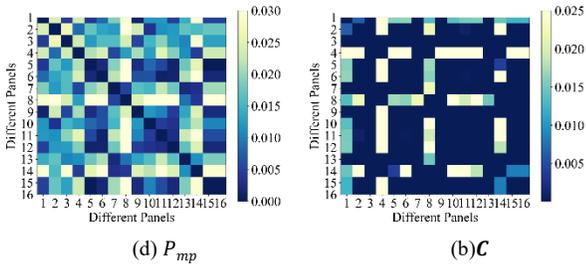

(d) $P_{mp}$          (b) $C$

Fig.4    The JS divergence for 16 PV modules(Case 4)

### B. Case study of degrading variables in common factor

As shown in (12), faults related variables, such as $C_o, O, V_{mp}$, are in common multiplier in the analytical expression of $C$. In this case study, $C_0$ of module 1 is set to be 0.822(0.8*optimal value); $V_{mp}$ of module 4 is set to be $0.8*V_{mpo}$; $C_o$ of module 8 is set to be 0.873(0.85*optimal value); $O$ of module 14 is set to be 0.85. The JS divergence between PDFs of power output and $C$ are shown in Fig.2.

It can be seen from Fig.2(a) that JS divergence between PDF of $P_{mp}$ of each module can hardly discriminate abnormal ones from the normal ones. Nevertheless, as is shown in Fig.2(b), JS divergence of PDF of $C$ between module 1 and the majority is large, so module 1 is considered to have a fault. It is consistent with our setting of module 1 as a fault one. The same situation happens to module 4,8,14. Hence, our method can correctly detect module abnormalities when soiling and shading condition, degradation faults or short-circuit faults occur.

Moreover, a greater JS divergence is detected between module 1(or module 4) and the majority than between module 8(or module 14) and the majority. It indicates that our method can reveal the severity of failures as well.

### C. Case study of degrading variables in $(A_1, A_2, A_3, A_4)$

In this case study, DHI and GHI of module 1 and 4 is set to be 0.85*normal value; DNI of module 8 and 14 is set to be 0.9*normal value. The JS divergence between PDFs of output power and $C$ are shown in Fig.3.

From Fig.3(a), it can be observed that JS divergence between PDF of $P_{mp}$ of each module can hardly discriminate abnormal ones from the normal ones. Nevertheless, as is shown in Fig.3(b), JS divergence of PDF of $C$ between module 1 and the majority is large, so module 1 is considered to have a fault. It is consistent with our setting of module 1 as a fault one. The same situation happens to module 4,8,14. Hence, our method can correctly detect module abnormalities when open-circuit faults occur.

### D. Case study of multiple faults

This case study intends to assess the ability of $C$ to detect the presence of multiple faults in modules. DHI and GHI of Module 1 is set to be 0.85; DNI of module 4 is set to be 0.85; $O$ of module 8 is set to be 0.85; $C_o$ of module 14 is set to be 0.85. The JS divergence between PDFs of output power and $C$ are shown in Fig.4.

As is illustrated in Fig.4(a), JS divergence between PDF of $P_{mp}$ of each module can hardly discriminate abnormal ones from the normal ones. Nevertheless, as is shown in Fig.4(b), JS divergence of PDF of $C$ between module 1 and the majority is large, so module 1 is considered to have a fault. It is consistent with our setting of module 1 as a fault one. The same situation happens to module 4,8,14. Thus, our method can correctly detect module abnormalities when various fault conditions in TABLE II occur at the same time.

## V. Conclusion

In this paper, a novel approach is proposed to detect abnormalities of PV modules, based on Sandia Array Performance Model and GMM. This approach is validated using simulated data acquired by explicit expressions for power at MPP in original SPAM and practical data of weather. The main contribution and future work is summarized as follows.

- The GMM technique is applied to power system uncertainty analysis and model stochastic power outputs of PV modules. In particular, a random vector $C$ is developed to eliminate differences of probability distribution of output power due to varying tilt and azimuth angles of PV modules. Therefore, the identification of faults is based on the comparison of probability density of $C$ instead of the power produced by different modules.
- This approach does not need the use of irradiance or temperature sensors as all other monitoring and failure

detection procedures do, which have been found in literatures,. This will reduce sensor investment in practice.
- Besides fault detection capability, the proposed approach is able to capture fault severity. The more severe the fault is, the greater JS divergence is. Therefore, it can be used to alert the user of the severity of failures and the priority to fix them.
- So far, the detection of faults is achieved provided that the power outputs at MPP, azimuth angles and tilt angles of PV modules are known. To identify the type/cause of failures, a more thorough analysis that also considers other electrical quantities - voltage and current should be undertaken as future work.

## REFERENCES

[1] Y. Wu, Q. Lan, and Y. Sun, "Application of BP neural network fault diagnosis in solar photovoltaic system," 2009 International Conference on Mechatronics and Automation, 2009.
[2] H. Mekki, A. Mellit and H. Salhi, "Artificial neural network-based modelling and fault detection of partial shaded photovoltaic modules", Simulation Modelling Practice and Theory, vol. 67, pp. 1-13, 2016.
[3] W. Rezgui, L.-H. Mouss, N. K. Mouss, M. D. Mouss, and M. Benbouzid, "A smart algorithm for the diagnosis of short-circuit faults in a photovoltaic generator," 2014 First International Conference on Green Energy ICGE 2014, 2014.
[4] K. Chao, P. Chen, M. Wang and C. Chen, "An Intelligent Fault Detection Method of a Photovoltaic Module Array Using Wireless Sensor Networks", International Journal of Distributed Sensor Networks, vol. 10, no. 5, p. 540147, 2014. Available: 10.1155/2014/540147.
[5] Y. Stauffer, D. Ferrario, E. Onillon, and A. Hutter, "Power monitoring based photovoltaic installation fault detection," 2015 International Conference on Renewable Energy Research and Applications (ICRERA), 2015.
[6] B. Kang, S. Kim, S. Bae and J. Park, "Diagnosis of Output Power Lowering in a PV Array by Using the Kalman-Filter Algorithm", IEEE Transactions on Energy Conversion, vol. 27, no. 4, pp. 885-894, 2012. Available: 10.1109/tec.2012.2217144.
[7] E. Garoudja, F. Harrou, Y. Sun, K. Kara, A. Chouder, and S. Silvestre, "A statistical-based approach for fault detection and diagnosis in a photovoltaic system," 2017 6th *International Conference on Systems and Control (ICSC)*, 2017.
[8] J. Duffie and W. Beckman, Solar energy thermal processes. New York: Wiley, 1974.
[9] King, D.; Boyson, W.; and Kratochvil, J. (2004). "Photovoltaic Array Performance Model." 41 pp.; Albuquerque, NM: Sandia National Laboratories. SAND2004-3535.
[10] Z. Chen, L. Wu, S. Cheng, P. Lin, Y. Wu, and W. Lin, "Intelligent fault diagnosis of photovoltaic arrays based on optimized kernel extreme learning machine and I-V characteristics," *Applied Energy*, vol. 204, pp. 912–931, 2017.
[11] R. Li, Z. Wang, C. Gu, F. Li and H. Wu, "A novel time-of-use tariff design based on Gaussian Mixture Model", Applied Energy, vol. 162, pp. 1530-1536, 2016.
[12] R. Singh, B. Pal, and R. Jabr, "Statistical Representation of Distribution System Loads Using Gaussian Mixture Model," IEEE Transactions on Power Systems, vol. 25, no. 1, pp. 29–37, 2010.
[13] G. Valverde, A. Saric and V. Terzija, "Probabilistic load flow with non-Gaussian correlated random variables using Gaussian mixture models", IET Generation, Transmission & Distribution, vol. 6, no. 7, p. 701, 2012.
[14] P.V.Mikheev,"Multidimensional Gaussian probability density and its applications in the degenerate case", Radiophysics and Quantum Electronics, vol.49, no.7, p.564-571, 2006.

**Qianni Cao** is an undergraduate at Wuhan University. She is going to pursue her Ph.D. degree in electrical engineering at Tsinghua University, Beijing, China. Her research interests include power system probabilistic analysis and renewable energy generation.

**Chen Shen** (M'98–SM'07) received his B.E. and Ph.D. degrees in Electrical Engineering from Tsinghua University, Beijing, China, in 1993 and 1998, respectively. From 1998 to 2001, he was a postdoc in the Department of Electrical Engineering and Computer Science at University of Missouri Rolla, MO, USA. From 2001 to 2002, he was a senior application developer with ISO New England Inc., MA, USA. He has been a Professor in the Department of Electrical Engineering at Tsinghua University since 2009. Currently, he is the Director of Research Center of Cloud Simulation and Intelligent Decisionmaking at Energy Internet Research Institute, Tsinghua University. He is the author/coauthor of more than 150 technical papers and 1 book, and holds 21 issued patents. His research interests include power system analysis and control, renewable energy generation and smart grids.

**Mengshuo Jia** (S'18) received his B.E. degree in Electrical Engineering from North China Electric Power University, BaoDing, China, in 2016. He is pursuing his Ph.D. degree in electrical engineering at Tsinghua University, Beijing, China. His research interests include power system probabilistic analysis and renewable energy generation.